\documentclass{article}

\PassOptionsToPackage{numbers, compress}{natbib}
\usepackage[nonatbib, final]{neurips_2024}

\usepackage[utf8]{inputenc}
\usepackage[T1]{fontenc}
\usepackage{xurl}
\usepackage[colorlinks, allcolors=mydarkblue]{hyperref}
\usepackage{booktabs}
\usepackage{amsfonts}
\usepackage{nicefrac}
\usepackage{microtype}
\usepackage{xcolor}
\usepackage{graphicx}
\usepackage{tabularx}
\usepackage{enumitem}
\usepackage{multirow}
\usepackage{array}
\usepackage{apacite}
\usepackage{cleveref}
\usepackage{float}
\let\citep\shortcite
\let\citet\shortciteA
\let\citealp\shortciteNP
\AtBeginDocument{\urlstyle{APACsame}}
\definecolor{mydarkblue}{rgb}{0,0.08,0.45}

\title{Third-party compliance reviews \\ for frontier AI safety frameworks}

\author{%
\parbox{0.8\linewidth}{\centering%
\mbox{Aidan Homewood\thanks{Corresponding author: \url{aidan.homewood@governance.ai}.}\hspace{1.15ex}$^{1}$\enskip} %
\mbox{Sophie Williams\hspace{0.2ex}$^{1}$\enskip} %
\mbox{Noemi Dreksler\hspace{0.2ex}$^{1}$\enskip} \rule[-1mm]{0pt}{7mm}%
\mbox{John Lidiard\hspace{0.2ex}$^{1}$\enskip} %
\mbox{Malcolm Murray\hspace{0.2ex}$^{2}$\enskip} %
\mbox{Lennart Heim\hspace{0.2ex}$^{1}$\enskip} %
\mbox{Marta Ziosi\thanks{The author contributed to this work in a personal capacity, independent of their role in the European Union General-Purpose AI Code of Practice.}\hspace{1.15ex}$^{3}$\enskip} \rule[-1mm]{0pt}{7mm}%
\mbox{Seán Ó hÉigeartaigh\hspace{0.2ex}$^{4}$\enskip} %
\mbox{Michael Chen\hspace{0.2ex}$^{5}$\enskip} %
\mbox{Kevin Wei\hspace{0.2ex}$^{6}$\enskip} \rule[-1mm]{0pt}{7mm}%
\mbox{Christoph Winter\hspace{0.2ex}$^{4,7}$\enskip} %
\mbox{Miles Brundage\hspace{0.2ex}$^{8}$\enskip} %
\mbox{Ben Garfinkel\hspace{0.2ex}$^{1}$\enskip} \rule[-1mm]{0pt}{7mm}%
\mbox{Jonas Schuett\hspace{0.2ex}$^{1}$\enskip}\\
}
\vspace{0.6em}
\\\\%
\small $^{1}$Centre for the Governance of AI
\hspace{0.5em}$^{2}$SaferAI
\hspace{0.5em}$^{3}$Oxford Martin AI Governance Initiative \\
\hspace{0.5em}$^{4}$Leverhulme Centre for the Future of Intelligence, University of Cambridge \\
\hspace{0.5em}$^{5}$METR
\hspace{0.5em}$^{6}$Harvard University
\hspace{0.5em}$^{7}$Institute for Law \& AI
\hspace{0.5em}$^{8}$Independent
}

\begin{document}
\maketitle
\setcounter{footnote}{0}

\begin{abstract}
Safety frameworks have emerged as a best practice for managing risks from frontier artificial intelligence (AI) systems. However, it may be difficult for stakeholders to know if companies are adhering to their frameworks. This paper explores a potential solution: third-party compliance reviews. During a third-party compliance review, an independent external party assesses whether a frontier AI company is complying with its safety framework. First, we discuss the main benefits and challenges of such reviews. On the one hand, they can increase compliance with safety frameworks and provide assurance to internal and external stakeholders. On the other hand, they can create information security risks, impose additional cost burdens, and cause reputational damage, but these challenges can be partially mitigated by drawing on best practices from other industries. Next, we answer practical questions about third-party compliance reviews, namely: (1) Who could conduct the review? (2) What information sources could the reviewer consider? (3) How could compliance with the safety framework be assessed? (4) What information about the review could be disclosed externally? (5) How could the findings guide development and deployment actions? (6) When could the reviews be conducted? For each question, we evaluate a set of plausible options. Finally, we suggest “minimalist”, “more ambitious”, and “comprehensive” approaches for each question that a frontier AI company could adopt.
\end{abstract}
\vspace{1.6em}

\newpage
\tableofcontents
\newpage

\section*{Executive summary}

This paper makes the case for third-party compliance reviews for frontier AI safety frameworks and answers practical questions about how to conduct them.

\subsection*{What are third-party compliance reviews? (\Cref{section1})}
During a third-party compliance review, an independent external party assesses whether a frontier AI company complies with its safety framework. Anthropic and G42 have already committed to commissioning such reviews, while the third draft of the EU General-Purpose AI Code of Practice recommends that companies assess whether they will adhere to their framework. Note that compliance reviews are distinct from adequacy reviews, which examine whether or not a safety framework and the way it is implemented are adequate for mitigating the risks posed by frontier AI systems.

\subsection*{The case for third-party compliance reviews (\Cref{section2})}

Third-party compliance reviews can benefit a frontier AI company in three main ways. First, they likely increase compliance with safety frameworks, which aim to keep risks associated with the development and deployment of frontier AI systems to an acceptable level. Second, they provide assurance to external stakeholders that the company is compliant with its safety framework (e.g. the public, government bodies, and other frontier AI companies). Third, they provide assurance to internal stakeholders (e.g. senior management, the board of directors, and employees).

However, third-party compliance reviews also present several challenges. For example, they can create new security risks if sensitive information is leaked. They can also impose substantial time and financial costs on frontier AI companies. Additionally, they could provide inaccurate results, leading to reputational damage or a false sense of security. Practical obstacles also arise, including measurability challenges and the risk of employee self-censorship. However, these challenges are not unique to frontier AI companies. They can be mitigated through measures often used in audit and assurance (e.g. segregating and monitoring reviewer duties, appointing an internal liaison, and choosing a competent reviewer).

\subsection*{How to conduct third-party compliance reviews (\Cref{section3})}

We identify six key aspects of compliance reviews and evaluate options for each of them.

\begin{figure}[H]
    \centering
    \includegraphics[width=\linewidth]{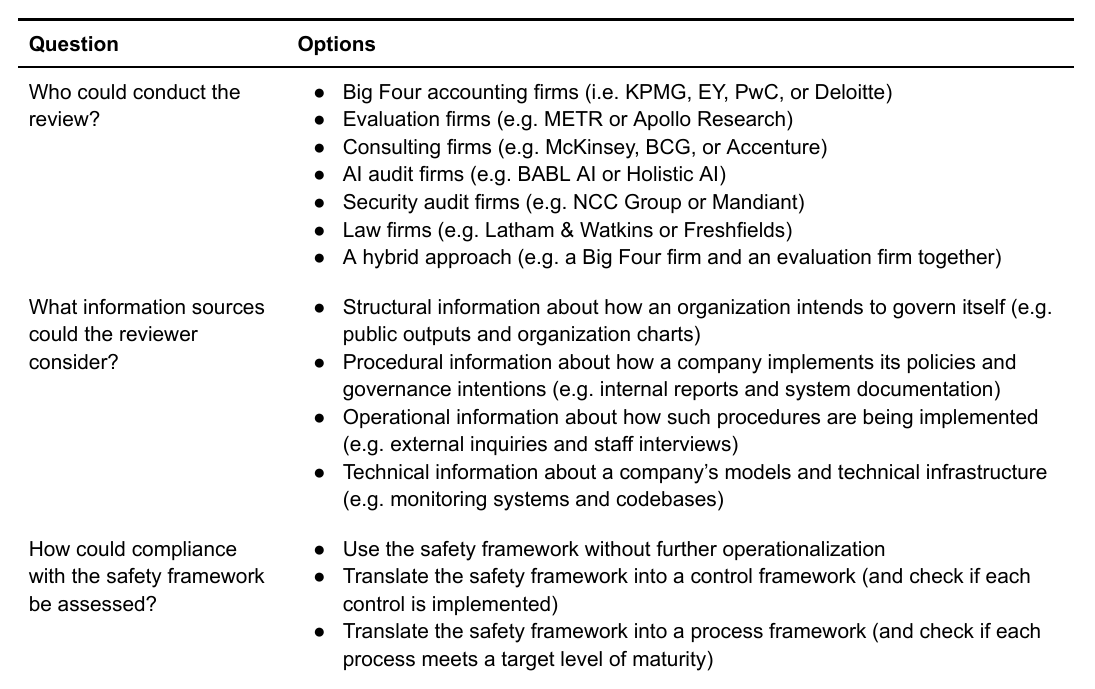}
\end{figure}

\newpage

\begin{figure}[H]
    \centering
    \includegraphics[width=\linewidth]{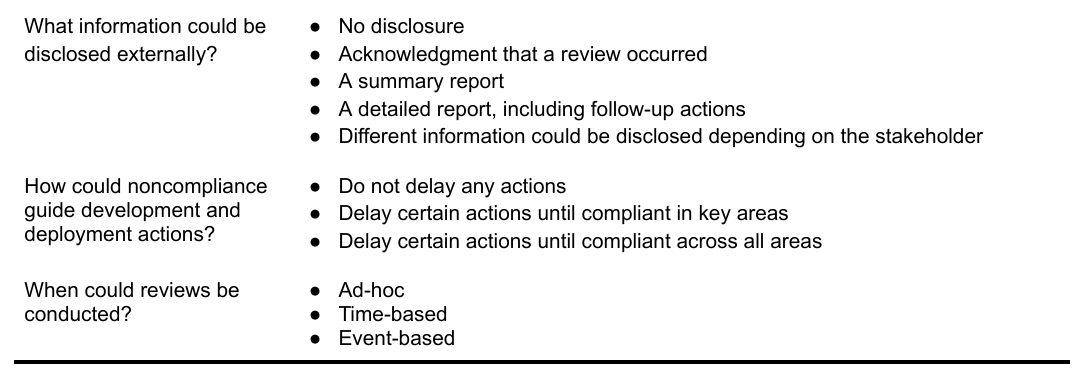}
\end{figure}

\subsection*{Suggesting different approaches (\Cref{section4})}

For each of the questions, we suggest a “minimalist”, “more ambitious”, and “comprehensive” approach that a frontier AI company could take.

\begin{figure}[H]
    \centering
    \includegraphics[width=\linewidth]{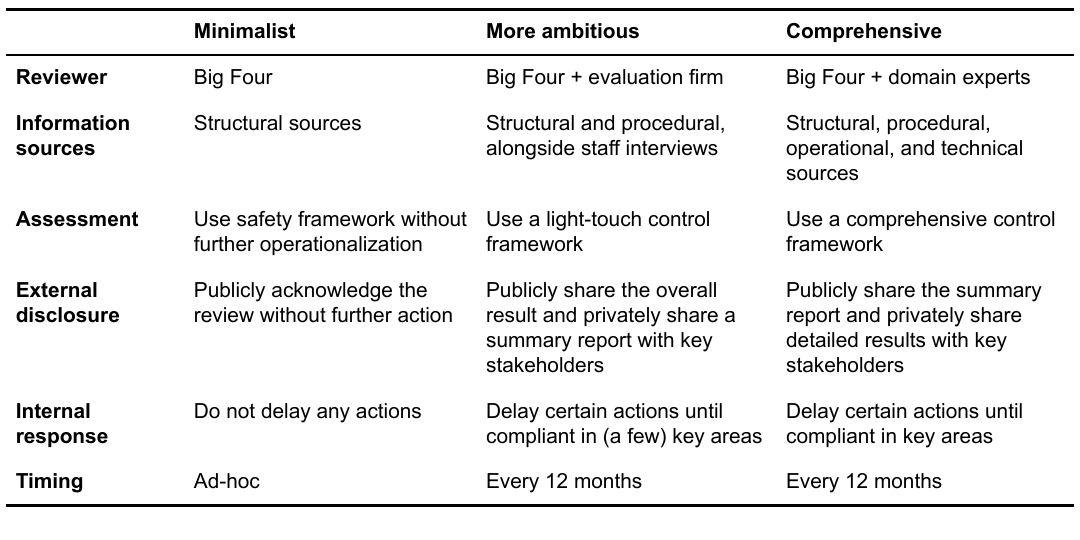}
\end{figure}

\newpage

\section{Introduction}\label{section1}

Frontier AI safety frameworks are a new type of risk management framework. Their main purpose is to keep risks associated with the development and deployment of frontier AI systems to an acceptable level. They typically focus on catastrophic risks (e.g. from chemical or biological weapons, large-scale cyberattacks, or loss of control). As of May 2025, 12 companies have published a safety framework \citep{metr2025}, including \citet{anthropic2025}, \citet{OpenAI2025-}, and \citet{googledeepmind2025}. They also play a key role in the Frontier AI Safety Commitments \citet{dsit2024} and the upcoming EU General-Purpose AI Code of Practice \citep{EuropeanCommission2025}. Safety frameworks have emerged as a best practice in frontier AI governance \citep{Buhl2025-,metr2025,FrontierModelForum2024}.

However, it often remains unclear whether frontier AI companies actually adhere to their safety frameworks. There is an inherent information asymmetry between the people who implement the framework and essentially all other stakeholders. This includes external stakeholders, such as customers, downstream AI developers, other frontier AI companies, the public, and governments. But it also includes internal stakeholders, such as senior management, the board of directors, and employees with limited visibility. As a result, internal and external stakeholders may not trust that companies are following their commitments.

This paper explores a potential solution to this problem: third-party compliance reviews. During a third-party compliance review, an independent external party assesses whether a company complies with an internal policy, standard, or law. Such reviews are common in other domains, such as financial services \citep{sarbanes_oxley_2002}, cybersecurity \citep{sabillon_2017}, nuclear power \citep{pang_2014},  oil and gas \citep{Short2016-}, and aviation \citep{low_2019}. The risks specific to frontier AI systems \citep{Anderljung2023} mean frontier AI companies are best suited to adopt third-party compliance reviews. In this paper, we focus on third-party reviews that assess whether a frontier AI company adheres to its own safety framework. 

Although there has been some practical work on this topic, scholarship remains limited. Most notably, \citet{anthropic2025} and \citet{G422025} have already made commitments to commission compliance reviews of their safety frameworks. Compliance reviews are also mentioned in the third draft of the EU General-Purpose AI Code of Practice \citep{EuropeanCommission2025}, while the Frontier AI Safety Commitments include reviews of internal accountability and governance \citep{dsit2024}. Third-party compliance reviews for frontier AI safety frameworks are recommended by several scholars \citep{alaga2024,Schoemaker2024,Campos2025-,Pistillo2025-}. However, there seems to be no practical guidance on how to conduct such reviews. Against this background, this paper seeks to answer the following research questions:

\begin{itemize}[leftmargin=2em]
    \item What are the benefits and challenges associated with third-party compliance reviews for frontier AI safety frameworks?
    \item How could third-party compliance reviews be conducted? In particular: (1) Who could conduct the review? (2) What information sources could the reviewer consider? (3) How could compliance with the framework be assessed? (4) What information could be disclosed externally? (5) How could the findings guide development and deployment actions? (6) When could reviews be conducted?
\end{itemize}

The paper has three areas of focus. First, it focuses on compliance reviews, as opposed to audits of AI systems \citep{Raji2020-,Falco2021-,Raji2022-outsider,Anderljung2023,Manheim2024-}, audits of data privacy \citep{Toy2017-,Liu2024-}, or reviews of the adequacy of frontier AI safety frameworks (Lidiard et al., forthcoming). Second, it focuses on third-party compliance reviews, as opposed to reviews conducted by internal teams \citep{Raji2020-,schuett2024-ia}. Third, it focuses on voluntary compliance reviews, as opposed to reviews that are required by law (though our analysis may also be relevant for mandatory reviews).

A note on terminology. In this paper, we talk about “reviews” instead of “audits”. The term “compliance review” typically refers to voluntary standards, often those that are imposed by an organization upon itself \citep{Coppersmith1997,Kempe2021-}. In contrast, “compliance audits” typically assess compliance with legal obligations. For example, the EU Digital Services Act mandates independent compliance audits for certain online platforms and search engines \citep{EuropeanCommission2023}. More generally, audits tend to involve more formal or standardized processes \citep{mokander_2021,mokander2023} and typically represent a higher level of assurance than reviews \citep{Badertscher2023-}.

This paper proceeds as follows. First, we make the case for third-party compliance reviews for frontier AI safety frameworks (\Cref{section2}). Next, we answer practical questions about how to conduct such reviews (\Cref{section3}). Finally, we suggest a “minimalist”, “more ambitious”, and “comprehensive” approach that a frontier AI company could take for each of the questions (\Cref{section4}). The paper concludes with a summary of our main contributions and suggestions for further research (\Cref{section5}).

\section{The case for third-party compliance reviews}\label{section2}

In this section, we make the case for third-party compliance reviews for frontier AI safety frameworks. We discuss some of the main benefits (\Cref{section2.1}) and challenges (\Cref{section2.2}).

\subsection{Benefits}\label{section2.1}

Third-party compliance reviews offer three key benefits to a frontier AI company.

\textbf{Increased compliance.} Third-party compliance reviews are likely to increase the extent to which companies adhere to their safety frameworks. For example, research has found that employees tend to improve their performance when anticipating an audit \citep{Neidermeyer2005-}. They also become more accountable to their peers \citep{Cardinaels2012-}, management \citep{Pilbeam2016-}, and auditors \citep{Ewelt-Knauer2021-}. Because reviews provide an independent perspective, they can also identify areas of noncompliance that companies themselves may overlook. Employees may even reveal compliance issues to reviewers that they had not shared with upper management \citep{Tynan2005-}. This can help to ensure that the objectives of the safety framework are met. 

\textbf{Assurance for external stakeholders.} Third-party compliance reviews provide assurance to external stakeholders through external validation. Internal teams could conduct similar reviews, but external stakeholders might question their reliability \citep{Brundage2020-,Avin2021-,Ashfaq2021-}. Independent reviews produce findings that stakeholders are more likely to view as trustworthy \citep{Ohman2012-}. Relevant stakeholders include government entities and society, as they may want to verify company claims about responsible AI development practices. Compliance reviews also provide other frontier AI companies with evidence that their competitors are implementing safety measures. This is particularly valuable given that some companies define acceptable risk relative to competitors’ deployments \citep{Meta2025-,OpenAI2025-}. Publication of the review’s results could also reduce pressure to release untested models \citep{Brundage2020-} thus lessening race dynamics. Compliance reviews may also provide assurance to downstream developers \citep{Williams2025-} and users, especially in cases where they have limited visibility into the foundation models they are using. This can enhance security confidence for customers who process sensitive data through these systems. As a result, a frontier AI company that proactively seeks compliance reviews may build a stronger reputation for themselves in the long term.

\textbf{Assurance for internal stakeholders.} Third-party compliance reviews also provide assurance to internal stakeholders about safety framework adherence. Internal stakeholders include senior management, board members, and employees working on unrelated projects. Generally, the company or reviewer will provide the board and management with a detailed report covering the review’s findings. This enables boards and management to effectively address safety framework compliance gaps. Other governing bodies such as Anthropic’s Long-Term Benefit Trust \citep{anthropic2023-longterm} or Google DeepMind’s AGI Safety Council \citep{dragan-2025} could receive a similar report. This would enable them to consider safety framework compliance when making decisions. Employees could receive lightly redacted reports under non-disclosure agreements, instead of the full version. This would still allow them to gain a detailed understanding of safety framework compliance and the reviewer’s recommendations, whilst presenting less risk than external disclosure.

\subsection{Challenges}\label{section2.2}

At the same time, third-party compliance reviews pose some challenges. Below, we set out four key challenges and suggest ways to address them.

\textbf{Security risks.} Third-party compliance reviews may increase security risks for a frontier AI company. Reviewers may have access to sensitive information about hardware and software providers, security systems, and internal security processes. This information could be used to identify security weaknesses \citep{nelson-2013,Thompson2017}. Furthermore, if a reviewer were to improperly store this information, this could create an attack surface for adversaries. Model weights represent one critical intellectual property at risk. Model weight theft threatens national competitiveness and security \citep{nevo2024}. Tacit information about the model development process is another critical intellectual property. This knowledge could similarly compromise national competitiveness and security. Third-party induced security incidents have been documented in other industries. One study found 247 such incidents in public firms between 2000 and 2020 \citep{Benaroch2021-}. In one notable example, an insider from Amazon Web Services identified a misconfiguration during a cloud security compliance review. This person later exploited this weakness to access Capital One’s cloud storage, exposing sensitive banking data. Although unlikely, this suggests reviewers could exploit sensitive information they encounter during reviews. This might result in intellectual property leakage or exposure of sensitive commercial information (e.g. model release schedules). Information leaks could also trigger negative media coverage.

A frontier AI company can mitigate security risks through techniques used on other highly sensitive industries. Most obviously, the company can also be selective about what information is shared with the reviewer (see \Cref{section3.2} and \Cref{section4}). Auditor activities (e.g. the files they read) can be tracked using an “audit trail” \citep{Layton2016,Duncan2016-,iso_2022}. Duties can be segregated and role-based access management can be implemented for individual review employees \citep{Hewitt2008-,aftab2019-,iso_2022}. Enhanced physical security can be established through document review rooms, asset controls (check in/check out), and escort protocols \citep{Lightman_2022}. For less sensitive material, secure document management and transfer software can be used \citep{Jagadeeswari2023-}. Companies can vet reviewers through processes similar to internal hiring procedures (see \citealp{iso_2022}).

Third-party compliance reviews may also improve security by identifying issues. This identification may allow for quick fixes or security process improvements. Private third parties routinely conduct security compliance reviews in industries with the highest security needs (e.g. banking and cloud services) \citep{Bryce_2019,fdic_2023,Lins_2018} . Anthropic already holds external certification for ISO 27001, an information security standard (see \citealp{AnthropicTrustCenter}), while OpenAI commits to third-party audits of their security controls \citep{OpenAI2025-}. The security benefits from compliance reviews may therefore outweigh the security risks they create.

\textbf{Cost burden.} Third-party compliance reviews will likely impose a cost burden on a frontier AI company. This could manifest in several ways, including the diversion of employee and management time, the use of internal resources, and direct monetary cost from reviewer fees. Negotiating the terms of the review and managing employees’ expectations requires time and attention from senior management. Employees may also be required to participate in interviews or engage in other time-consuming interactions with reviewers. Legal teams, in particular, might need to oversee document disclosure and monitor reviewer access to sensitive information, especially if there are concerns about confidentiality breaches or other legal risks \citep{Berkowitz2021}. Reviewer fees typically scale with the time spent on the engagement, meaning longer or more complex reviews can drive up costs. If relevant information is fragmented across multiple teams or departments this may result in a prolonged review process. Some of these costs do not vary with the scope of the review, so third-party compliance reviews may be too costly for smaller AI companies with limited budgets and management capacity.

However, several strategies drawn from auditing and assurance practices can help to manage costs. First, a frontier AI company could adopt a more minimalist approach to the review (see \Cref{section4}). Second, appointing a dedicated review liaison can streamline communications and coordination \citep{Ismael2018-}. Third, early engagement with legal teams can proactively address potential issues related to confidentiality and access, reducing bottlenecks later in the process. Fourth, initiating a structured kick-off session between reviewers and company employees can help set clear expectations and promote efficiency from the outset. Finally, the frontier AI company could complete an internal compliance review in advance, allowing them to identify and resolve issues before the third-party review begins \citep{Abbott2012-}.

\textbf{False results.} Third-party compliance reviews may provide false results that mislead a frontier AI company and their stakeholders. False positives occur when reviewers deem a company compliant with its safety framework when it is not. This creates a false sense of security within the company and among stakeholders. Reviewers face strong incentives that can compromise their objectivity. Auditors want to retain large clients, particularly in concentrated markets, which may lead to overly positive reviews \citep{Wills2025}. Smaller auditors could face even stronger retention pressure since large clients are likely to represent a larger portion of their overall revenue. Reviewers may have a conflict of interest if they provide the company with other services (e.g. consulting or model evaluations), which could lead to false positive results \citep{Salehi2009-,Tepalagul2015-}. However, firms typically seek highly reputable auditors \citep{Barton2010-}, which also provides auditors with an incentive to maintain their reputation. False negatives present a different problem. They occur when reviewers incorrectly determine that a company is not compliant with its safety framework. This can subject companies to unjustified negative attention. 

A frontier AI company can reduce the risk of false results through several approaches. First, they can ensure reviewers receive adequate information to complete effective reviews (or, at the very least, acknowledge where information might be insufficient). Second, they can ensure that reviewers possess the expertise and resources needed to conduct an effective review – for example, by opting to employ multiple reviewers (see \Cref{section3.1}). Third, they can reduce potential conflicts of interest – for example, by choosing a reviewer not providing other services for a fee (e.g. consulting, model evaluations).

\textbf{Measurability challenges.} Measuring compliance with the commitments in the safety framework also presents significant challenges. Key commitments may be subjective or open to interpretation, potentially setting a low bar for certifying a frontier AI company as safe. In some cases, thresholds for commitments evolve through an unspecified process \citep{Kasirzadeh2024-}, creating ambiguity when evaluating if a threshold was met and whether corresponding mitigations were applied. OpenAI's initial Preparedness Framework illustrates this problem by stating that risk thresholds (low, medium, high, and critical) are “speculative and will keep being refined” \citep{OpenAI2023-tt}. Without internal documentation providing additional clarity, significant time may be needed to define or operationalize aspects of the safety framework. Alternatively, reviewers might declare in their report that they cannot determine compliance with a requirement due to insufficient information. Compliance reviews lose effectiveness without sufficient operationalization, potentially becoming mere box-ticking exercises for commitments so vague they cover a wide range of company behaviors.

A frontier AI company can improve measurability through several approaches. One approach is to enhance future iterations of safety frameworks or internal documentation. For example, companies could clarify the intended scope and frequency of compliance reviews in their safety framework. Companies could also enhance future safety frameworks by replacing optional clauses with if-then commitments \citep{Karnofsky2024}. For example, instead of writing that they “may also solicit external expert input prior to making final decisions on the capability and safeguards assessments” \citep{anthropic2025}, they could specify conditions that would trigger expert consultation. Companies could also specify their response to new information or practices. For example, rather than stating that they “expect concrete measures implemented to reach each level of security to evolve substantially” \citep{GoogleDeepMind2025-2}, they could specify the sources they expect to rely on for new information (e.g. security reports from leading think tanks), establish methods for deciding which measures merit implementation, and assign responsibility for implementing new measures. Another way to improve measurability would be to require reviewers to highlight any commitments requiring further operationalization after completion of a review. 

\textbf{Self-censorship.} Employee self-censorship also presents a significant limitation of third-party compliance reviews. Employees may voluntarily withhold or alter information they provide to reviewers due to three main factors: psychological concerns, organizational influences, and legal considerations. Psychological factors emerge when employees are concerned about retaliation from peers or management for speaking up \citep{Ewelt-Knauer2021-,Lee2021-}. This can create reporting gaps \citep{obrien2024} or chilling effects throughout the organization. Employees may also believe their disclosures will not lead to meaningful change, discouraging them from revealing safety framework compliance concerns \citep{Lee2021-}. Organizational factors include employees’ desire to maintain high financial compensation from their employer \citep{Balafoutas2020-}. The company culture may also foster a “code of silence” where employees subtly or explicitly agree to remain silent about compliance concerns. For example, one study found that employees may act dishonestly as a group when collective rewards exist \citep{Balafoutas2020-}. This is particularly important if model deployment, or increases to the market capitalization of a frontier AI company, act as collective rewards. Finally, legal considerations can also shape employee self-censorship, since frontier AI companies may be within scope of several authorities in the United States. This includes the Defense Production Act, the Export Administration Regulations, the International Emergency Economic Powers Act, and the Federal Trade Commission consumer protection authorities \citep{Bullock2024}. Reviews that reveal awareness of potential harms without appropriate responses may establish negligence in subsequent legal proceedings, increasing company liability. In addition, employees with confidentiality agreements might worry about contract breaches when sharing information with external reviewers.

Companies can mitigate employee self-censorship through a range of interventions. Organizations can implement anonymous reporting systems or surveys that allow employees to share concerns without fear of identification. They can conduct private interviews designed to be credibly anonymous and build trust with employees. Companies can establish robust whistleblower protections that shield employees who report compliance concerns \citep{Anderljung2023}. Management can clearly explain to employees their exact legal obligations, while clarifying the scope of confidentiality agreements. They can also set explicit expectations about what information employees should share with reviewers. Finally, companies can offer financial bonuses for compliance and apply penalties for noncompliance \citep{Fochmann2020}.

The limitations described above may raise concerns among a frontier AI company considering third-party compliance reviews. However, auditing practices in other industries demonstrate that it may be possible to mitigate them effectively. 

\section{Key aspects of the review}\label{section3}

In this section, we examine six aspects of third-party compliance reviews for frontier AI safety frameworks, namely who could conduct the review (\Cref{section3.1}), what information sources they could consider (\Cref{section3.2}), what assessment framework could be used (\Cref{section3.3}), what information could be disclosed externally (\Cref{section3.4}), how AI companies could respond to findings (\Cref{section3.5}), and when a review could be completed (\Cref{section3.6}). For each question, we list plausible options and discuss their main advantages and disadvantages.

\subsection{Who could conduct the review?}\label{section3.1}

We assess six options for who could conduct compliance reviews: Big Four accounting firms, evaluation firms, consulting firms, AI audit firms, security audit firms, and law firms. Our assessment is based on five criteria: (1) AI expertise, (2) risk management expertise, (3) compliance review expertise, (4) public reputation, and (5) ability to handle sensitive information (see \Cref{table3}).

\textbf{Big Four firms.} One set of potential reviewers are the Big Four accounting firms: KPMG, EY, PwC and Deloitte. These four professional services firms conduct financial audits for virtually all large companies in the world. In 2017, all 500 companies in the S\&P 500 index used a Big Four auditor \citep{Gow_2018}. They also conduct compliance audits with legal requirements like the EU Digital Services Act, including securities regulation \citep{Peterson_2017} and Environmental, Social and Governance internal frameworks and internal standards \citep{Gipper2024-}. Big Four firms offer several benefits as compliance reviewers. They may have superior risk management expertise from their work in other industries, which may cause a greater reduction in corporate risk compared to other audit firms \citep{Bley2018-}. They have compliance audit and compliance review expertise across different industries, including banks and defense clients \citep{Gow_2018}. This expertise suggests their superior security resources and experience \citep{Rosati2020-}. Collusion and fraud are less likely than with other categories of firms \citep{Lennox_2008}, though their professional standards may vary depending on the context. Big Four reviewers provide greater assurance to stakeholders because of their good reputation \citep{Ege2025}. However, these firms also face limitations. They may have very limited frontier AI expertise. Although each firm has set up responsible AI teams \citep{Bruno2024}, they do not usually focus on frontier AI. The improved quality of Big Four audits may rely on partner quality \citep{Che2019-}, and experienced partners may not yet be available in the AI governance context. It remains unclear if they would find this market attractive; the segment might be too small relative to their established business, and they might perceive potential reputational risks.

\textbf{AI evaluation firms.} AI evaluation firms, such as METR and Apollo Research, represent another set of potential reviewers. These specialized organizations focus on technical assessments of AI systems, particularly frontier models. They offer technical evaluation services ranging from model testing and benchmarking to vulnerability assessments and safety analyses for frontier AI companies. AI evaluation firms offer important benefits. Their extensive frontier AI expertise places them at the frontier of evaluation science. Existing relationships with frontier AI companies suggest that these firms have already addressed security concerns and internal cost limitations. Such established connections may also foster trust from other frontier AI companies in the review results. But these firms also have possible limitations as compliance reviewers. Despite having potential for growth with proper investment, these firms currently maintain limited organizational capacity. Compliance reviews represent uncharted territory, as these companies have no previous experience conducting such assessments. The narrow focus on AI also prevents these evaluators from applying relevant insights from similar compliance work in other industries.

\textbf{Consulting firms.} Consulting firms, particularly McKinsey, BCG, Bain \& Company, Accenture, and Capgemini represent another set of potential reviewers. These professional services firms provide strategic and operational advice on business activities across diverse sectors and contexts. Their services typically focus on improving organizational performance, strategy formulation, and implementation of business solutions. Consulting firms offer several benefits for compliance reviews. Their experience across industries may provide valuable risk management expertise. These firms maintain robust protocols for appropriately handling sensitive information, demonstrated through their work with government and banking clients. Their established reputations in the business community likely provide greater assurance to stakeholders regarding the credibility of review results. But these firms also have limitations in the context of a safety framework compliance review. Unlike accounting firms, most consulting organizations do not typically offer structured compliance review services as part of their core offerings. Although they possess well-developed security resources and experience through engagements with large companies, banks, and national security clients, their capabilities mirror those of Big Four firms without additional advantages. Much like the Big Four, their frontier AI expertise remains limited. While some consulting firms have established specialized AI teams, such as Bain \& Company's Responsible AI program, these initiatives rarely focus on frontier AI technologies and their unique governance challenges.

\begin{table}[t]
    \makebox[\textwidth][c]{\includegraphics[width=\linewidth]{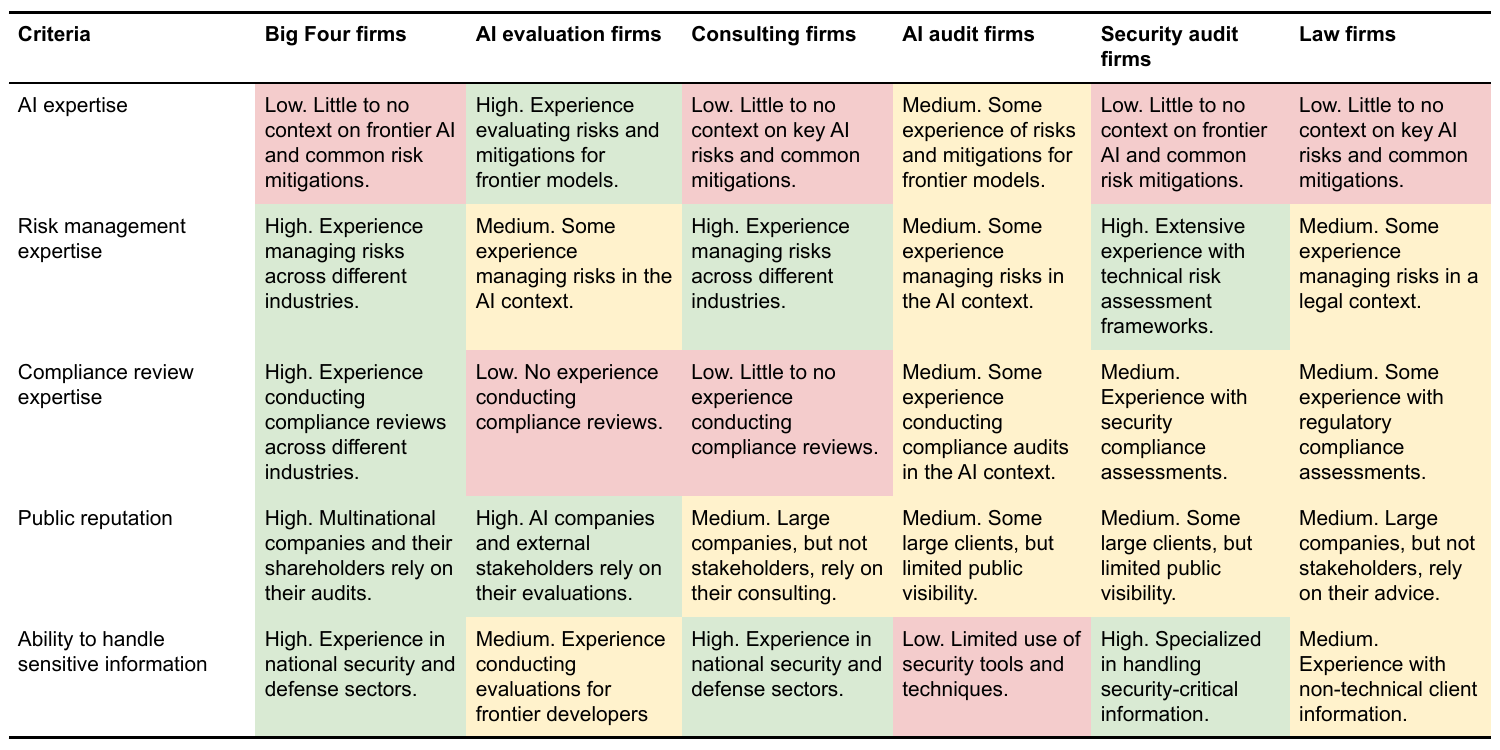}}
    \vspace{-0.5em}
    \caption{Comparative analysis of potential reviewers.}
    \label{table3}
\end{table}

\textbf{AI audit firms.} Specialized AI audit firms, such as BABL AI and Holistic AI, are another set of potential reviewers. These boutique organizations offer dedicated AI audit services focused on compliance with standards like ISO 42001, HIPAA, and the NYC Bias Audit law. However, their expertise typically centers on downstream applications rather than foundation models, which may limit their ability to address the unique challenges of frontier AI systems. AI audit firms provide several distinctive benefits. Some possess frontier AI expertise along with a comprehensive understanding of evaluation methodologies and widely implemented safety measures. Their familiarity with organizational policies in the AI context enables them to navigate internal governance structures effectively. Several of these firms demonstrate particular skill in translating abstract internal policies into concrete operational processes that support compliance objectives. But limitations also constrain the suitability of AI audit firms. Their specialized focus on AI means they lack insights into similar compliance work in other regulated industries. Their security capabilities present another concern, as these organizations typically maintain limited security resources and experience. This deficiency stems from the relatively lower security requirements of their usual clients – downstream AI deployers – compared to the robust security needs of banks, defense contractors, and national security clients.

\textbf{Security audit firms.} Security audit firms, such as NCC Group, IOActive, and Mandiant (now part of Google), represent another potential set of reviewers. These specialized companies evaluate organizations' cybersecurity practices, information systems, and technical infrastructure for vulnerabilities and compliance with security standards. They typically conduct penetration testing, vulnerability assessments, and security compliance reviews across various industries. Security audit firms offer several benefits. They possess strong technical expertise in evaluating complex systems and identifying security weaknesses. Many security audit firms also have experience handling highly sensitive information and maintaining confidentiality. Their familiarity with risk assessment frameworks and various compliance standards provides valuable context for evaluating compliance with frontier AI safety frameworks. However, these firms also have limitations. Most security audit firms have limited AI-specific expertise, particularly in regard to frontier AI capabilities and risks. Their assessments typically focus on technical security concerns rather than broader governance issues relevant to AI compliance. Furthermore, while security audit firms understand technical controls, they may have less experience with the organizational governance structures necessary for effective frontier AI risk management.

\textbf{Law firms.} The final set of potential reviewers we identify is law firms. Big law firms like Latham \& Watkins or Freshfields regularly advise on legal compliance and conduct internal investigations, and sometimes serve as third-party monitors to ensure ongoing regulatory adherence \citep{Martinez2013}. A big law firm with a dedicated technology practice, or a specialized AI governance firm (e.g. ZwillGen), may be particularly appropriate for conducting AI compliance reviews. Law firms offer several unique benefits for compliance reviews. Many have established expertise in conducting internal investigations \citep{Davis2019}. Their professionals possess refined skills in gathering evidence through various methods, particularly interviews with organizational stakeholders. A significant advantage in many jurisdictions is that attorney-client privilege can provide legal protection for sensitive communications and findings generated during the review process. But law firms also have limitations in the AI compliance context. Even law firms specializing in AI governance likely lack the frontier AI expertise necessary for evaluating advanced systems. Most top legal practices have minimal experience conducting comprehensive compliance reviews compared to their traditional advisory roles. Their security capabilities present another concern, as these firms typically maintain limited security resources and technical experience. While they routinely handle legally sensitive information, they rarely interact with sensitive details about IT and cybersecurity systems. Finally, law firms generally lack experience operationalizing internal policy.

\begin{figure}[t]
    \makebox[\textwidth][c]{\includegraphics[width=0.6\linewidth]{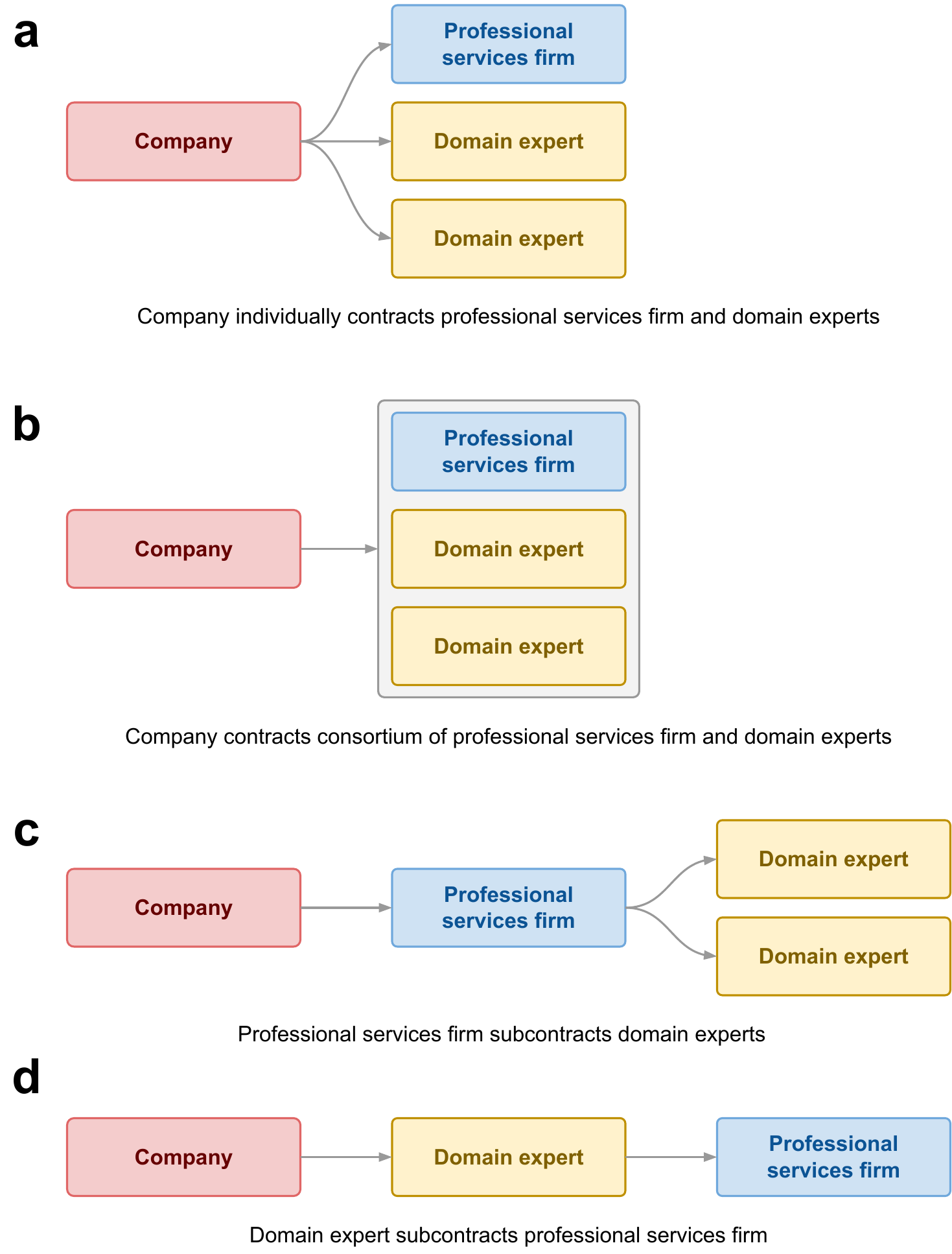}}
    \vspace{-0.5em}
    \caption{Different models for potential hybrid approaches.}
    \label{figure1}
\end{figure}

\textbf{Hybrid approaches.} Two or more of the above firms could collectively carry out a compliance review. This collaborative approach could be implemented through several structures: contracting different firms individually, engaging a single consortium of providers, or contracting a primary firm that then subcontracts one or more additional firms (see Figure 1). Big Four accounting firms routinely engage domain experts as subcontractors to complete specialized audits or assurance engagements. Although less common, they also occasionally serve as subcontractors themselves, sometimes loaning professional staff to support the internal audit functions of client companies \citep{Glover2006}. Hybrid approaches offer significant benefits for complex compliance reviews. The primary advantage is complementary expertise – a gap in one firm's capabilities can be effectively addressed by partnering with another organization that possesses the required competencies. For example, combining a Big Four accounting firm with an AI evaluation firm could create a comprehensive review team with both established compliance methodologies and frontier AI expertise, while maintaining a high ability to handle sensitive information (see \Cref{table3}). However, hybrid approaches introduce several limitations. A frontier AI company may face higher setup costs due to increased contractual complexity when multiple service providers must be engaged simultaneously. This challenge could potentially be mitigated by working with a Big Four firm that maintains existing subcontractor relationships or internal capabilities for identifying and managing specialized partners. Organizational clarity presents another concern – contracting different firms individually or engaging a consortium may create confusion among employees and relevant stakeholders about roles and responsibilities. Finally, these approaches may result in blurred accountability when multiple organizations share responsibility for review outcomes, potentially complicating remediation efforts if deficiencies are identified.

Our analysis demonstrates distinct strengths and limitations across the various potential reviewers. Each category offers unique capabilities that may be valuable depending on specific organizational needs and compliance contexts. The comparative assessment highlights important tradeoffs between technical AI expertise, knowledge of risk management, established compliance methodologies, reputation, and security capabilities. Organizations considering compliance reviews must carefully weigh these factors against their particular requirements. Hybrid approaches that combine complementary capabilities from different reviewer categories may provide comprehensive solutions that address the complex challenges of frontier AI safety framework compliance.

\subsection{What information sources could the reviewer consider?}\label{section3.2}

Reviewers can use various information sources to assess a company's adherence to its safety framework. We classify these sources into four tiers: (1) structural, (2) procedural, (3) operational, and (4) technical. \Cref{table4} provides a list of information sources with ratings to indicate their confidentiality.

\begin{table}[t]
    \makebox[\textwidth][c]{\includegraphics[width=0.8\linewidth]{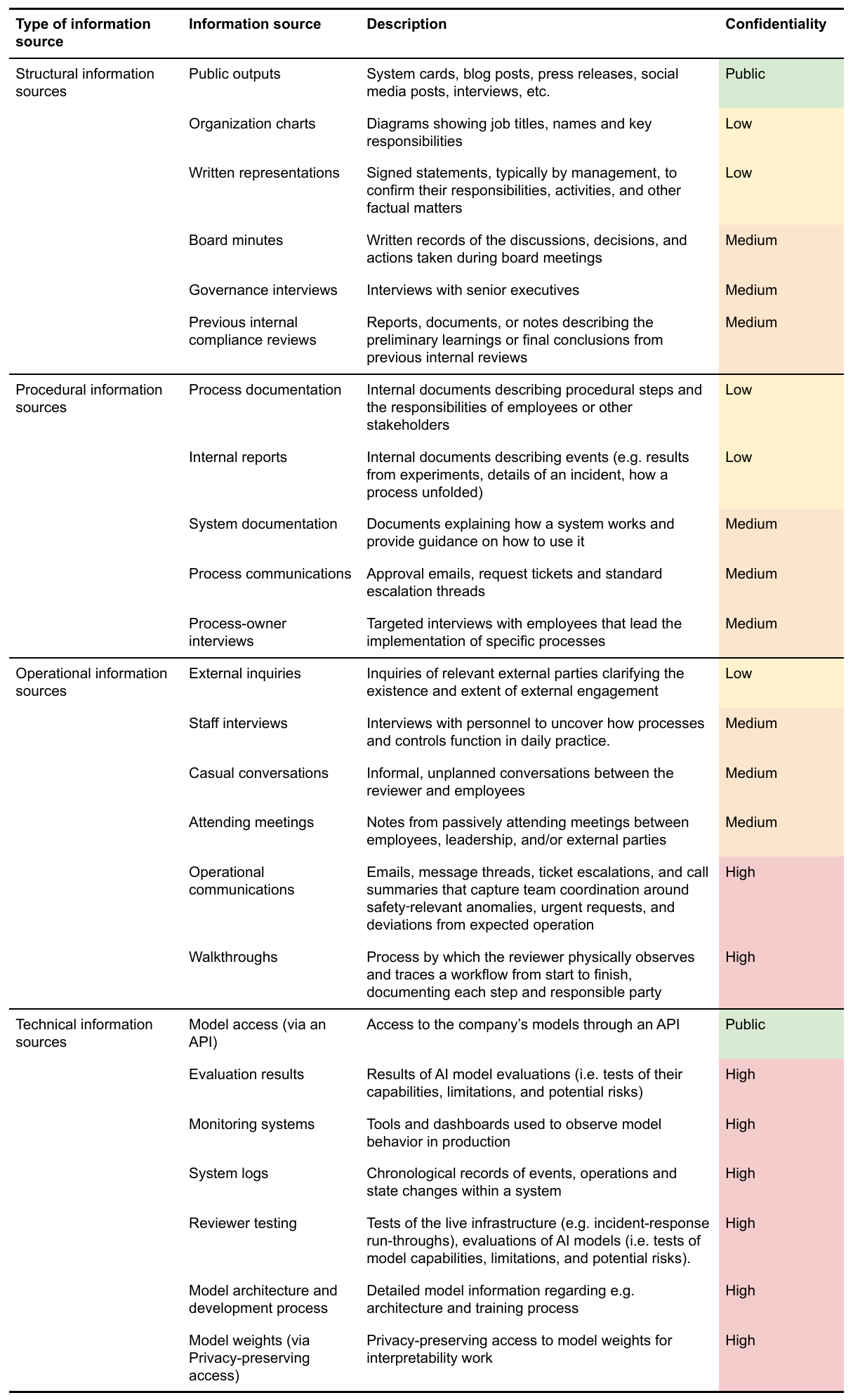}}
    \vspace{-0.5em}
    \caption{List of information sources and their level of confidentiality.}
    \label{table4}
\end{table}

\textbf{Structural information sources.} Structural information sources provide high‑level information about how an organization intends to govern safety. These sources include organizational charts, board minutes, and public outputs. These sources allow reviewers to map formal lines of responsibility and risk management commitments. Many industries commonly use these sources in audits. Structural information sources offer some key benefits. First, requests for information can be handled by senior staff and kept to a minimum. Second, these sources consist mainly of electronic documents, meaning that costs are kept low. Third, information security risks remain low as reviewers do not need access to production systems or sensitive data. However, restricting a review to these structural information sources also has limitations. First, reviewers may not be able to identify compliance gaps or behavioral drift because the information is high-level. As a result, the findings would provide limited assurance of actual compliance. Second, key indicators of operational failures or control weaknesses (i.e. logs, internal communications, or process documentation) are unavailable, limiting the potential for corroboration between sources. Third, these sources do not provide insight into the safety culture or staff experiences. Fourth, using these sources alone may lead to recommendations that are too generic to lead to meaningful improvements. 

\textbf{Procedural information sources.} Procedural information sources reveal how a company implements its policies and governance intentions. These include system documentation, logs, and process communications. These sources allow reviewers to confirm that controls are in place, trace specific events, and uncover procedural weaknesses that structural sources may miss. These sources are routinely used in audits, particularly within the technology sector. Procedural information sources provide additional benefits beyond structural ones. First, they allow reviewers to assess if relevant processes and controls exist. Second, reviewers can make more confident findings, providing stakeholders with greater assurance. However, limitations exist with procedural information sources too. For example, access to logs or ticketing systems could surface personal data, model details, or trade secrets. Risk reduction methods (e.g. redaction, masking, on‑premise viewing, and audit‑trail logging) reduce this risk but add coordination time and effort. Furthermore, reviewers are unable to factor in organizational culture, tacit knowledge, and undocumented norms. Finally, staff might self‑censor or move sensitive discussions to untraceable channels, reducing the value of examining their communications (see \Cref{section2.2}).

\textbf{Operational information sources}. Operational information sources provide direct or near-real‑time visibility into granular tasks, decision‑making, and team interactions. These sources include staff interviews, operational communications, and walkthroughs. Their purpose is to verify that controls function as intended while surfacing tacit knowledge and behavioral drift. These sources are routinely used in audits, particularly within the technology sector. Operational information sources offer further benefits beyond structural and procedural ones. These sources allow reviewers to uncover tacit, undocumented information about company activities. These sources can also allow reviewers to make more confident findings, providing stakeholders with greater assurance. In addition, these sources allow reviewers to make more practical recommendations about how to improve compliance. However, tacit information about security, alongside reviewer testing, could further exacerbate information security risks. Various teams would need to provide considerable input to organize observations, prepare test environments, and manage external inquiries. 

\textbf{Technical information sources.} Technical information sources originate from the deployed model and its supporting infrastructure. These include API access to AI models, evaluation results, and model architecture and development process. Reviewers can use these sources to verify the implementation of technical requirements and safeguards mandated by the safety framework. Technical information sources offer a number of benefits. These sources allow reviewers to test safety claims with actual system data rather than relying solely on internal reports or system cards. Second, these sources can contain fine-grained technical evidence to identify the causes of safety failures. This allows reviewers to produce confident findings, providing the highest level of stakeholder assurance. Finally, reviewers could better scrutinize the development process with more access to technical information, which could improve the credibility of the review \citep{Casper_2024}. However, reviewer access to technical information sources also has limitations. Access to model internals can risk intellectual property theft, data leakage, and potential misuse. In addition, many system-level tests require specialized expertise and significant computing resources which auditors may not have. Furthermore, to the extent that these benefits require access to model weights, additional progress in privacy-preserving technology is necessary to facilitate safe access \citep{bucknall_2023}.

In summary, a more expansive set of information sources strengthens a review by increasing the breadth and granularity of evidence. This allows reviewers to make precise and realistic findings and recommendations, which may improve compliance. However, access to broader and more granular information increases security and privacy risks, and raises coordination challenges. This creates trade-offs in deciding what information sources to use. The next section explores frameworks for assessing information sources to assess compliance.

\subsection{How could compliance with the safety framework be assessed?}\label{section3.3}

Before a reviewer can check whether a company is doing what its safety framework states, they must decide how they will translate that framework into something they can measure. This operationalization\footnote{Turning broad or abstract commitments into check-points that can be scored is called operationalization and makes them observable and testable in practice.} process first requires identifying the commitments that a frontier AI company has made in its safety framework. These commitments can be individual intentions, policies, actions, and methods for managing risks from frontier AI systems \citep{alaga2024,metr2025,Buhl2025-} and can vary in their level of abstraction. A reviewer can then assess compliance with the safety framework commitments by (1) grading the safety framework commitments without further operationalization, (2) operationalizing the safety framework commitments into a control framework, or (3) operationalizing the safety framework commitments into a process maturity framework. 

\textbf{Assess without further operationalization.} The reviewer could provide a grade for each of the commitments in the safety framework (e.g. “solicit internal and external feedback on the capabilities report”) without operationalizing them. The grade for each commitment could be on a simple pass/fail scale \citep{Iso2015-}, a degree of compliance (e.g. “compliant”, “partially-compliant”, and “noncompliant”), or a degree of confidence (e.g. from “not at all confident” to “fully confident” that the company is compliant), although the last is not commonly used. There are two benefits of this option. First, the results will be easy to understand, particularly for external stakeholders, as they map directly to the safety framework. Second, this approach is less costly because it does not require an internal team or an external firm to further operationalize the safety framework. However, there are limitations to this approach. First, compliance with the commitments may be difficult to measure, and so the assessment of compliance could be highly open to interpretation. Second, this approach limits the ability for a frontier AI company to improve its processes, which may reduce compliance in the long term.

\begin{table}[t]
    \makebox[\textwidth][c]{\includegraphics[width=0.8\linewidth]{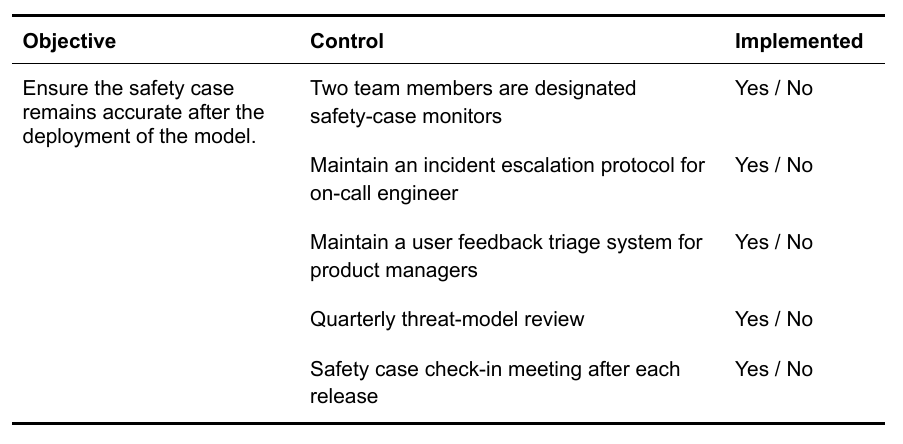}}
    \vspace{-0.5em}
    \caption{Compliance grading rubric for a single mock objective and five mock controls.}
    \label{table5}
\end{table}

\textbf{Operationalize into a control framework.} Alternatively, the company could use a control framework to operationalize the commitments in the safety framework. Controls are policies, processes, or technical measures that modify risk \citep{iso_2018}. The first step would be to list their objectives – the concrete outcomes that must hold for each commitment in their safety framework to be achieved (e.g. “the safety case remains accurate after deployment”).\footnote{Objectives can span multiple aspects and levels \citep{iso_2018}. Similarly, commitments in a safety framework can be at different levels of abstraction, so may require varying levels of operationalization to be turned into objectives.} Before a compliance review, a frontier AI company – or a contracted professional services firm – would identify controls that are already in place or design new ones. \Cref{table5} illustrates how a single objective can be supported by several controls. The designed controls would need to be implemented before the review. During the review, the assessor tests a random set of controls for each objective, and notes any deviations and uncertainty in their report. Control-based assurance is common in sectors such as finance \citep{sarbanes_oxley_2002} and technology \citep{AICPA2017}, and is already being used by frontier AI companies for other purposes (see e.g. \citealp{AnthropicTrustCenter}). This approach has a number of benefits. First, controls are easier to follow compared to vague safety framework requirements, and might therefore lead to greater compliance. Second, the approach provides better guidance on addressing areas of noncompliance, as individual commitments can be targeted. Third, control frameworks are common, meaning the company can draw on existing standards \citep{COSO2013-,iso_2018} and expertise. However, there are limitations to this approach. For example, it could have a high cost burden, particularly if a professional services firm is required. In addition, if they decide to repeat the review process, the company may need to change the control framework. This approach could also reduce the company’s flexibility in response to emerging risks, if the controls are too rigid and difficult to adapt.

\textbf{Operationalize into a process maturity framework.} Another way of operationalizing the commitments in the safety framework is using a process maturity framework. This type of framework allows a reviewer to assess the maturity of a company's processes, even if those processes are complex or undefined \citep{iia-2013}. Process maturity frameworks are commonly used in software development \citep{Helgesson2011-}, particularly for US federal government technology projects \citep{Mishra_2016,dod-2024}. While process maturity frameworks can be adapted to assess compliance \citep{schrimpf_2020}, they are not commonly used for this purpose. Before conducting the compliance review, a frontier AI company – or a contracted professional services firm – would create a framework for grading processes.\footnote{Alternatively, an existing process maturity framework can be adapted. One widely used framework is Capability Maturity Model Integration (CMMI) \citep{cmmi-model} which distinguishes between five levels: (1) initial, (2) managed, (3) defined, (4) quantitatively managed, and (5) optimizing. Other popular frameworks include the Process Maturity Framework (PMF) \citep{Humphrey1998-}, Capability Maturity Model (CMM) \citep{Bate1994-}, Business Process Maturity Model (BPMM) \citep{ObjectManagementGroup2008-}, and Process and Enterprise Maturity Model (PEMM) \citep{Hammer2017-}.} The first step would be to list the processes required to implement the safety framework commitments. Next, the expectations for each process would be described (see \Cref{table6} as an example). Finally, the target level for each process would be specified. Often this target is the third level, “defined \& managed”, as not all processes need to operate at the highest level of maturity \citep{iia-2013}. During the review, the reviewer would use the process maturity framework and information sources to evaluate the current level of maturity, and compare this to the target level of maturity. Upon meeting the target level for a process, the company would be compliant with the corresponding commitment or commitments. Using a process maturity framework has a number of benefits. First, processes are typically easier to follow compared to vague safety framework requirements, which might lead to greater compliance. Second, it allows reviewers to highlight the individual components of processes to be targeted for improvement in their recommendations, also likely improving compliance. Third, it encourages the company to continuously improve their processes outside of third-party compliance reviews. However, this approach also has limitations. First, creating and implementing a process maturity framework is likely a large cost burden to the frontier AI company. Second, to the extent processes must be continuously improved to be fully compliant, this approach would continue to be burdensome. Third, this approach risks overemphasizing process improvements rather than point-in-time improvements (i.e. a single fix that does not need ongoing work).

The assessment approach used changes what is assessed, how the results are presented, and the specificity of recommendations. How these results could be shared externally is discussed in the next section.

\subsection{What information could be disclosed externally?}\label{section3.4}
After completing a review, companies may share information with external stakeholders. Companies have four disclosure options: (1) no information about the review, (2) a statement that the review occurred, (3) a summary report, and (4) a detailed report including planned follow-up actions. Companies may choose to share different information depending on the stakeholder, which could include the public, enterprise customers, industry bodies, and relevant government entities.\footnote{We have chosen to focus on these stakeholders for brevity, but note that there may be others to consider in practice (e.g. regulators, insurers, and investors).}

\begin{table}[t]
    \makebox[\textwidth][c]{\includegraphics[width=0.9\linewidth]{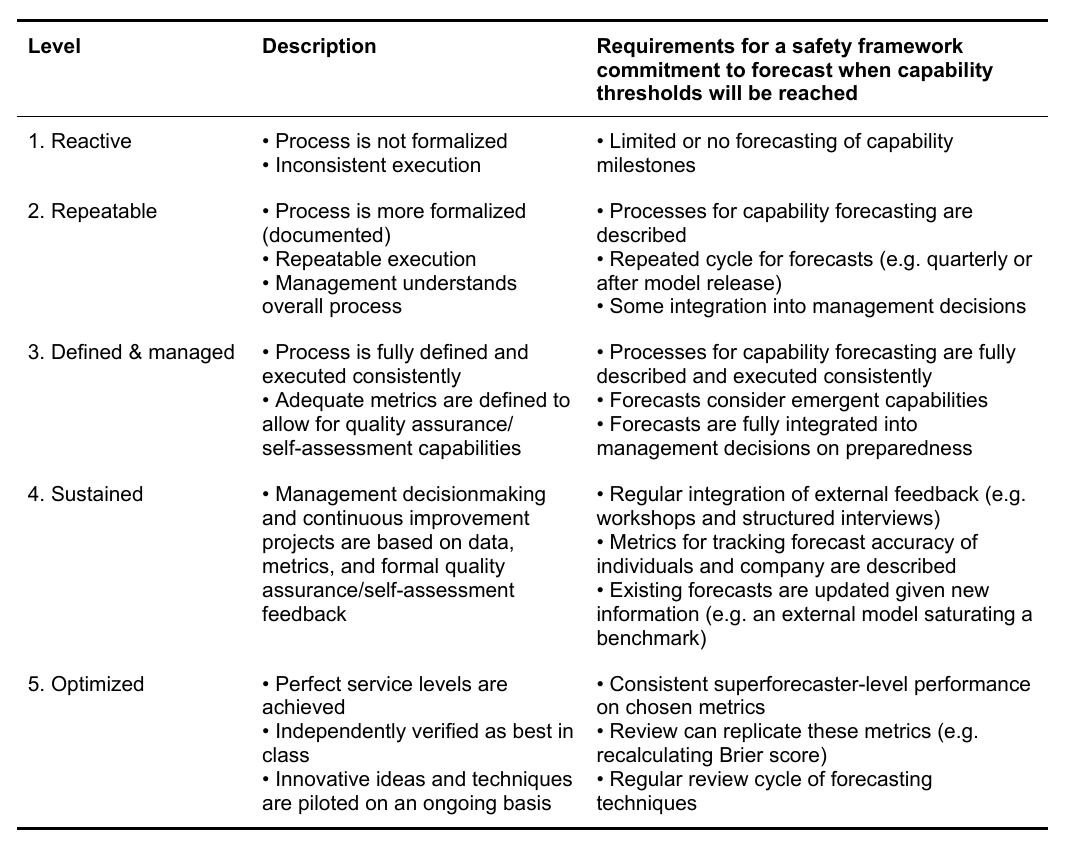}}
    \vspace{-0.5em}
    \caption{%
        Generic maturity framework \protect\citep{iia-2013} alongside a specific example for a commitment to forecast when capability thresholds will be reached.}
    \label{table6}
\end{table}

\textbf{No information about the review.} A frontier AI company may choose not to disclose that a compliance review occurred or to release any of the findings. Keeping compliance reviews confidential reduces reputational risks, and may make employees and management less averse to reviewer activities. This can lead to more open and collaborative reviews. Keeping the reviewers private also reduces the potential that they become attack surfaces for security threats. However, this approach provides no assurance to external stakeholders, which is a key benefit of conducting reviews. Companies would also miss out on potential reputational gains from a review.

\textbf{A statement that the review occurred.} A frontier AI company could publicly acknowledge that a review took place, potentially including the overall result, but without sharing material information about it. This might take the form of a blog post or social media announcement. The review process, findings, and follow-up actions would remain a private and internal matter. This approach mirrors practices in the technology sector, where companies advertise their certification for a standard (e.g. SOC 2, ISO 27001) while keeping audit reports confidential except under non-disclosure agreements. In the nuclear energy industry, WANO review interactions and results are confidential \citep{Kovynev2014-}, but a nuclear power plant can disclose that a review took place (e.g. \citealp{TVO2012-}). A key benefit to this approach is that the company can reap some reputational benefits while avoiding the risks associated with more transparent approaches (e.g. security risks). However, this approach only provides minimal assurance to external stakeholders, and without external accountability, companies may lack remediation incentives. The limited information about the review might also generate public speculation and follow-up questions from government bodies. 

\textbf{A summary report.} A frontier AI company could publish a summary report that presents high-level information about the review and its key takeaways. The report could be tailored to suit the needs of different stakeholders. To minimize biases, the summary could be authored by the reviewer, as is often done in the financial sector \citep{Liu2020-}. Publishing such a report would provide external stakeholders with greater assurance than a simple statement that a review took place. It could also increase the company’s motivation to address identified shortcomings. Additionally, sharing the summary with industry bodies could facilitate peer learning across the sector. On the other hand, negative findings might damage the company’s reputation. Stakeholders could also be concerned about “cherry-picking” or the omission of negative findings. One way to mitigate these risks would be to explicitly outline the context, methodology, and any caveats associated with the report. 

\textbf{A detailed report including planned follow-up actions.} A frontier AI company could publish a detailed report describing the review process, findings, and planned follow-up actions. Confidential information could be redacted. This report could be authored, or at least formally signed off, by the reviewer. This approach would strengthen transparency and accountability, creating strong incentives for compliance \citep{mokander2023}. It would also provide additional assurance to external stakeholders, allowing those with specific interests (e.g. data privacy and security) to reference relevant information. Moreover, it may be particularly effective in facilitating sector-wide learning. However, stakeholders might suspect that redactions are arbitrary or designed to obscure unfavorable findings. Furthermore, companies may be reluctant to disclose this level of detail, especially if the results are negative or reveal proprietary information, and lengthy internal approval processes could delay transparency efforts. Additionally, companies may have legitimate information security concerns and so choose to limit disclosure to select stakeholders (e.g. enterprise customers or government bodies) under non-disclosure agreements, similar to what Anthropic and OpenAI already do \citep{AnthropicTrustCenter,openai-trust-portal}. Sharing detailed results with government actors could enhance state capacity and deepen insight into frontier AI company activities, with existing partnerships and cybersecurity infrastructure reducing security risks. If other stakeholders received less detailed information, governments could offer assurances about the reliability of reviews, similar to financial sector disclosure frameworks and bank stability examinations \citep{FDIC2005-}. Nevertheless, detailed disclosure to regulatory-adjacent bodies might create chilling effects, potentially leading to company self-censorship. 

External disclosure of information from a compliance review serves as a key accountability mechanism that helps build trust amongst stakeholders. More extensive disclosure can enhance their reputation by demonstrating a commitment to transparency and strengthening their dedication to improved compliance. At the same time, broad disclosure of detailed findings raises legitimate privacy and security concerns. To balance these trade-offs, companies must carefully consider which stakeholders receive what information.

\subsection{How could noncompliance guide development and deployment actions?}\label{section3.5}

In the event the reviewer documents gaps between the stated policy and actual practice, a frontier AI company can respond in a number of ways. One important consideration is whether the company delays certain development and deployment actions (e.g. deploying a model internally, deploying a model publicly, releasing model weights publicly, training models, activating new capabilities) if they are found to be noncompliant. The company could (1) not delay any actions, (2) delay certain actions until they are compliant in key areas, or (3) delay certain actions until they are compliant across all areas. Different internal response options will lead to different levels of safety improvement while presenting different costs to the organization.

\textbf{Do not delay any actions.} The company could choose not to delay any development and deployment actions as a result of the review findings. This approach provides benefits, such as offering more operational flexibility and minimizing disruption as companies continue their activities regardless of the findings. This option may therefore be appealing for carrying out a first review. However, this approach also has limitations. For example, it does not create strong incentives to address findings, potentially reducing compliance. It also fails to provide additional assurance to stakeholders that the company will comply with its safety framework.

\textbf{Delay certain actions until compliant in key areas}. A company could delay certain development and deployment actions until they are compliant in some “key areas”. These key areas could include processes that are important for maintaining deployment mitigations. The company could decide where they are not compliant by selecting findings from the results of the review that they agree with. Before a review, the company could define two elements: (1) the actions they will delay if found to be noncompliant in key areas, and (2) the key areas for compliance (e.g. the process of determining whether a capability threshold has been reached). This approach provides a number of benefits. First, it creates an incentive to address noncompliance in key areas. It also provides greater assurance to stakeholders that the company is complying with its safety framework. However, this approach has various limitations. Importantly, it may result in inadequate compliance fixes being implemented by the frontier AI company, given their incentive to proceed with development and deployment actions. This approach may also increase internal costs and decrease buy-in for compliance reviews, as employees may be frustrated with the degree of compliance needed to continue with development and deployment activities.

\textbf{Delay certain actions until compliant across all areas.} A frontier AI company could delay certain development and deployment actions until they are compliant across all areas of the compliance review. This would require the company to define the actions that would be delayed if they were found to be noncompliant before the review is carried out. However, the company may first want to ensure that they agree with the findings of the review before delaying certain actions. There are benefits to this approach. Importantly, it creates the greatest incentive to address all areas of noncompliance identified by the review. It also provides the greatest practicable level of assurance for stakeholders with access to information about the response. But the approach also has limitations. Actions to address compliance issues may again be inadequate if the company is incentivized to address them as quickly as possible. This approach could significantly increase the cost burden for companies and decrease internal buy-in for compliance reviews. Furthermore, it severely limits flexibility for the company, as compliance is required across all areas of the review. 

Delaying certain actions until a frontier AI company achieves some level of compliance with the safety framework could improve assurance to stakeholders and increase compliance overall. However, delaying certain actions could limit flexibility and exacerbate the financial costs associated with a review.

\subsection{When could the reviews be conducted?}\label{section3.6}

A single review cannot ensure long-term compliance with safety frameworks. Review frequency is therefore important. The frequency of reviews can be: (1) ad-hoc, (2) time-based, or (3) event-based.

\textbf{Ad-hoc.} Companies can commission compliance reviews without fixed schedules. This approach may be appealing to companies in the early stages of adopting compliance reviews. Ad-hoc reviews allow companies to align review timing with internal capacity and organizational bandwidth. However, ad-hoc scheduling may create long gaps between review periods and delay detection of any issues. Furthermore, stakeholders only receive limited assurance from this approach because they cannot be confident that reviews will continue, and companies might only schedule reviews when compliance indicators appear favorable. 

\textbf{Time-based.} Time-based reviews occur at regular intervals (e.g. annually or semi-annually) which companies set in advance. Annual reviews represent the standard in many industries. Time-based reviews ensure regular attention to safety framework compliance. In addition, stakeholders have greater assurance that companies will maintain long-term compliance if they are aware of review schedules. However, reviews scheduled too infrequently may prove insufficient given rapid changes in frontier AI development. Conversely, overly frequent reviews may cause organizational fatigue and create unjustifiable costs. Regular schedules might also trigger “review-time behavior” where employees temporarily alter practices around expected review periods.

\textbf{Event-based.} Event-based reviews would occur in response to predefined triggers, such as planned model releases or major incidents. This approach aligns reviews with periods of heightened risk or potential impact, turning “review-time behavior” into a potential advantage. It also ensures scrutiny at critical junctures without requiring frequent routine reviews. However, reviews tied to a model’s release could result in its release being delayed, which companies are likely to resist. Furthermore, long intervals between trigger events could allow compliance issues to develop undetected.

In summary, review frequency is an important factor for ensuring their effectiveness over the long term. Overly frequent reviews may dilute quality and strain resources, while infrequent reviews risk missing important compliance issues. Therefore, it is important to strike a balance.

\section{Suggesting different approaches}\label{section4}

In this section, we identify the trade-off companies face for each aspect of the review. We then suggest “minimalist”, “more ambitious”, and “comprehensive” approaches that a frontier AI company could take for each aspect of the review. \Cref{table7} provides an overview of the different approaches for each aspect. Companies can mix different approaches or use options not covered in this analysis.

\textbf{Reviewer.} The selection of the reviewer types presents distinct expertise, reputation, and cost considerations for frontier AI companies. In \Cref{section3.1}, we discussed six potential reviewers: Big Four accounting firms, evaluation firms, consulting firms, AI audit firms, security audit firms, and law firms. When selecting reviewers, a frontier AI company faces a trade-off between the breadth of expertise the reviewer has, and the financial burden of the review. If the company prioritizes reducing their financial burden over ensuring the review team has technical expertise, they may want to only engage a Big Four accounting firm without additional subcontractors (minimalist approach). This is because Big Four firms demonstrate advantages across all approaches due to their superior capability in handling sensitive information, public reputation, extensive risk management expertise, and established compliance review experience. However, Big Four firms do not have much AI expertise. Therefore, if the company prioritizes technical expertise and financial cost similarly, they may want to engage a Big Four accounting firm with an evaluation firm as a subcontractor (more ambitious approach). If the company prioritizes technical expertise over financial cost, they may want to engage a Big Four accounting firm with multiple specialized subcontractors covering different domains (e.g. evaluation firms, cybersecurity specialists) to address all relevant technical aspects of AI safety (comprehensive approach). Other combinations, such as a security audit firm subcontracting an AI evaluation firm, could also provide enough expertise though they may lack compliance review experience. 

\textbf{Information sources.} The selection of information sources determines both the confidence in review findings and potential security exposure. In \Cref{section3.2}, we discussed four different types of information sources: structural (i.e. high-level governance intentions and actions), procedural (i.e. intended procedures), operational (i.e. implementation of procedures), and technical (i.e. models and technical infrastructure). A frontier AI company faces a trade-off between enabling confident reviewer findings and minimizing information security risks. A reviewer may not be able to recommend useful improvements to processes without access to procedural information. Therefore, if the company prioritizes information security risks over confident findings or recommendations, they may want to rely solely on structural sources (minimalist approach). If they weigh confident findings or recommendations and these information security risks similarly, they may want to use structural and procedural sources, as well as a selection of operational ones, such as staff interviews (more ambitious approach). If they prioritize confident findings or recommendations over these information security risks, or if they can mitigate these risks, they may want to use all four types of information sources (comprehensive approach). Companies may exercise discretion by exempting certain sources that they view as particularly sensitive or costly for the reviewer to acquire.

\begin{table}[t]
    \centering
    \includegraphics[width=\linewidth]{Figures/table2.pdf}
    \vspace{-1.5em}
    \caption{Minimalist, more ambitious, and comprehensive options for third-party compliance reviews.}
    \label{table7}
\end{table}

\textbf{Assessment.} Different forms of assessment offer varying levels of specificity and operational clarity, but present financial costs. In \Cref{section3.3}, we discussed three possible assessment framework options: assess without further operationalization, use a control framework, and use a process maturity framework. A frontier AI company faces a trade-off between gaining specific findings and investing time in authoring a control framework. If the frontier AI company prioritizes reducing upfront costs over improving specificity, they may want to simply use the safety framework directly for assessment (minimalist approach). If they weigh improving specificity and reducing upfront costs similarly, they may want to implement a light-touch control framework that specifies key commitments (more ambitious approach). If they prioritize improving specificity over reducing upfront costs, they may want to develop a more detailed control framework that maps objectives to an extensive set of controls (comprehensive approach). Process maturity frameworks likely go beyond the scope of a compliance review, as they require the company to define multiple maturity levels, not just the single level that corresponds to compliance.

\textbf{External disclosure.} External disclosure decisions shape stakeholder trust while determining reputational and information security risks. In \Cref{section3.4}, we discussed four options for what companies could disclose: no information, a statement that the review occurred, a summary report, or detailed results. A frontier AI company faces a trade-off between providing stakeholders with greater assurance and exposing themselves to information security risks when deciding what to disclose. We also discussed different stakeholders with whom companies may want to share information, including the public, enterprise customers, industry bodies, and relevant government bodies. More selective disclosure to these stakeholders can help navigate this trade-off, but may make stakeholders suspicious that important issues were hidden. If the company prioritizes information security risks over providing greater assurance to stakeholders, they may want to simply publicly acknowledge that a review occurred, using their discretion to decide whether to disclose the result (minimalist approach). If the company weighs providing greater assurance and these information security risks similarly, they may want to publish the overall result of the review, while only sharing the summary report with enterprise customers, industry bodies, and relevant government bodies (more ambitious approach). If the company prioritizes providing greater assurance over these information security risks, or if they can mitigate these risks, they may want to publish a summary report, and share a more complete version with enterprise customers, industry bodies, and relevant government bodies under appropriate confidentiality agreements (comprehensive approach).

\textbf{Internal response.} A company could implement various restrictions on development and deployment actions if they are found to be noncompliant by the review. In \Cref{section3.5}, we discussed three options for how companies could respond to review findings: do not delay any actions, delay certain actions until compliant in key areas, or delay certain actions until compliant across all areas. The frontier AI company faces a trade-off between creating stronger compliance incentives and maintaining operational flexibility when deciding how to respond to review findings. If they prioritize flexibility over compliance, they may not want to commit to delaying any actions if found to be noncompliant (minimalist approach). If they weigh compliance and flexibility similarly, they may want to commit to delaying certain actions until they are compliant in a few key areas (more ambitious approach). If they prioritize compliance over flexibility, they may want to commit to delaying certain actions until they are compliant in a broader set of key areas (comprehensive approach).

\textbf{Review timing.} The scheduling of compliance reviews affects both ongoing compliance and the cost burden a company faces. In \Cref{section3.6}, we discussed three options for when to conduct a review: ad-hoc, time-driven, and event-driven. The frontier AI company faces a trade-off between maintaining continuous compliance and minimizing ongoing costs when deciding when to conduct reviews. Event-driven reviews might balance assurance and cost by targeting high-risk periods, but may become burdensome, particularly for companies operating multiple models. If the frontier AI company prioritizes cost reduction over continuous compliance, they may want to conduct reviews on an ad-hoc basis (minimalist approach). If they weigh maintaining continuous compliance and avoiding ongoing costs similarly, they may want to undertake annual reviews (more ambitious approach). If they prioritize maintaining continuous compliance over avoiding ongoing costs, they may still favor annual reviews, as the additional benefits may not outweigh the additional cost burden (comprehensive approach).

\textbf{Interdependencies.} Different components of compliance reviews depend on each other. While companies may vary in ambition across different aspects of the review, choices in one aspect can shape or constrain what is feasible in another. For example, if a company limits the review to structural information sources, the value of subcontracting an evaluation firm diminishes. This is because a Big Four firm can assess this information without involving domain experts. Similarly, if using structural information sources alone, it may be inappropriate to use a control framework. This is because a control framework typically requires more granular information to assess compliance. Furthermore, the longer-term ambitions of a company may need to influence early design decisions. For example, a company aiming for a comprehensive control framework may benefit from adopting a “light-touch” control framework initially. This approach creates a foundation to build upon rather than starting with a type of framework that does not align with the end goal. Companies can mix and match approaches but options need to remain compatible.

\textbf{Alternative solutions.} Alternative solutions to third-party compliance reviews include internal compliance reviews and compliance management systems. Internal compliance reviews involve an internal but independent team carrying out the reviewer role \citep{schuett2024-ia}. Internal reviewers may better understand the company than third-party ones. This review could be completed before a third-party review to reduce internal costs \citep{Abbott2012-}. However, internal reviewers may not be able to provide the risk management expertise and independence that a third-party reviewer would provide. This means that internal and third-party reviews may serve as complements rather than substitutes. Compliance management systems promote compliance through dedicated staff (e.g. a Responsible Scaling Officer), monitoring and evaluation (e.g. compliance reviews), training and communication (e.g. escalation protocol awareness), and management structures (e.g. reporting hotlines, disciplinary measures) \citep{coglianese2021,doj-2024}. A compliance management system could be created after a third-party review using a professional services firm or by hiring employees with compliance management system expertise. Given third-party compliance reviews can form a part of a compliance management system, they are also complements rather than substitutes.

Selecting an approach to third-party compliance reviews requires careful evaluation of the trade-offs between potential benefits and risks. Companies do not need to apply the same level of ambition across all areas. Instead, they can adopt targeted strategies aligned with their specific priorities and risk profiles. Nevertheless, we think that either the minimalist and more ambitious approaches discussed are within reach for leading frontier AI companies undertaking their first third-party compliance review. As the industry matures, companies can progressively adopt more comprehensive approaches, allowing them to build compliance capacity while effectively managing risks. 

\section{Conclusion}\label{section5}
This paper has made the case for third-party compliance reviews for frontier AI safety frameworks, discussed different options for conducting such reviews, and provided guidance on how companies could decide what approach to take. 

The benefits of third-party compliance reviews include increased compliance with safety frameworks and assurance to both internal and external stakeholders. Despite these benefits, compliance reviews may create security risks if sensitive information is leaked, damage a company's reputation or create a false sense of security if results are incorrect, impose costs on the organization, and face the practical challenges of operationalizing and grading subjective commitments and employee self-censorship. Nonetheless, we believe that these can be mitigated using established audit and assurance practices common in other industries.

Our analysis addresses six key questions about how to conduct third-party compliance reviews: who could conduct the review, what information sources could be considered, how compliance could be assessed, what information could be disclosed externally, how noncompliance could guide development and deployment decisions, and when the reviews could be conducted. For each question, we describe minimalist, more ambitious, and comprehensive approaches that reflect different trade-offs between the costs and benefits. Each organization can calibrate its efforts to match its ambition and risk appetite. Over time, companies can build toward more comprehensive approaches by implementing mitigations for the challenges identified.

The paper left several questions unanswered that warrant further research. In particular, more research and experimentation are needed on which organizations or combinations of organizations are best positioned to conduct third-party compliance reviews for frontier AI safety frameworks, as the unique technical complexities and novel risks of these systems create significant reviewer selection challenges. Future work could examine how process maturity frameworks or control frameworks could be used to make safety framework requirements more concrete in practice. An examination of equivalent regulatory approaches would also be valuable to understand how governments could mandate compliance audits for frontier AI safety frameworks. Further work may also be needed on internal compliance reviews, which could be carried out at a higher frequency and improve internal buy-in compared to third-party compliance reviews. Additional work to develop compliance management systems for AI safety frameworks may also prove useful, particularly if frontier AI companies begin commissioning third-party compliance reviews. Finally, future research may be needed to help standardize results from third-party compliance reviews across the industry. Of course, if companies were to commission compliance reviews, their experience could shape further research directions.

The AI governance field is not alone in its challenge to gain insight into company activities – established audit and assurance best practices from other industries offer a strong foundation to build upon. Implementing third-party compliance reviews can help ensure that frontier AI companies remain aligned with their safety frameworks, fostering more robust risk management and strengthening stakeholder trust. By embracing these practices, companies not only improve their compliance but also demonstrate a commitment to responsible innovation. Through proactive investment in third-party reviews, frontier AI companies can better prepare for future regulatory requirements and demonstrate leadership in frontier AI governance. 

\section*{Acknowledgments}\label{acknowledgments}

We are grateful for valuable comments and feedback from 
Andreas Drechsler, Connor Stewart Hunter, Edward Kembery, Jared Perlo, Krzysztof Bar, Rajiv Dattani, Simon Mylius, Vael Gates, Vinay Hiremath, Zaheed Kara, and the participants of a workshop on auditing frontier AI safety frameworks co-organized by FAR.AI in Berkeley. All remaining errors are our own.

\bibliographystyle{apacite}
\bibliography{ms}

\end{document}